\newcommand{\m}{\rm \,m}
\newcommand{\mm}{\rm \,mm}

\newcommand{\sr}{\rm \,sr}

\newcommand{\GeV}{\rm \,GeV}

\newcommand{\numu}{$\nu_{\mu}$ }

\newcommand{\nutau}{$\nu_{\tau}$ }

\newcommand{\lsim}{\lower .5ex\hbox{$\buildrel < \over {\sim}$}}
\newcommand{\gsim}{\lower .5ex\hbox{$\buildrel > \over {\sim}$}}

\documentclass[epj]{svjour}
\usepackage{graphics}
\begin{document}
\title{MACRO results on atmospheric neutrino oscillations}
%\headnote{see Ref. [1] for a list of MACRO Authors    and  Institutions}
\author{G. Giacomelli\inst{1} \and A. Margiotta\inst{1} \\For
  the MACRO Collaboration \\
  \email{giacomelli@bo.infn.it} \and \email{margiotta@bo.infn.it}% etc
}                     % Do not remove
\institute{Dipartimento di Fisica and INFN, I-40127 Bologna, Italy \\
Paper presented at the HEP EPS Conf., Aachen, Germany, July 2003}
\date { }
\abstract{
The final results of the MACRO experiment on atmospheric neutrino
oscillations are presented. 
The data concern different event topologies with average neutrino energies
of $\sim 3$ and $\sim 50$ GeV. Multiple Coulomb Scattering of the high
energy muons was used to estimate the neutrino energy event by event. The angular distributions, the $L/E_\nu$ distribution, the particle
ratios and the absolute fluxes  all favour $\nu_\mu \rightarrow \nu_\tau$
oscillations with maximal mixing and $\Delta m^2 \simeq 0.0023 \: \rm eV^2$. 
Emphasis is given to measured ratios which are not affected by Monte Carlo (MC) absolute normalization; a discussion  is made on MC uncertainties.
 % end of PACS codes
} %end of abstract
\maketitle
\section{Introduction}
\label{intro}
MACRO was a large area multipurpose underground detector \cite{r1} designed
to search for rare events and rare phenomena in the penetrating cosmic
radiation.  It was located in Hall B of  the underground Gran Sasso Lab at
an average rock  overburden of 3700 m.w.e.; it
started data taking with part of the apparatus in \( 1989 \); it was
completed in \( 1995 \) and  was running in its final configuration until
the end of 2000. 
The detector had global dimensions of \( 76.6\times12 \times9 .3{\m }^{3}
\) and provided a total acceptance to an isotropic flux of particles of \(
\sim 10,000{\m }^{2}{\sr } \); vertically it was divided into a lower part,
which contained 10 horizontal layers of streamer tubes, 7 of rock absorbers
and 2 layers of liquid  scintillators, and an upper part which contained the
electronics and was  covered by 1 layer of scintillators and 4 layers of
streamer tubes. The sides were covered with 1 vertical layer of
scintillators and 6 of limited streamer tubes.\\
MACRO detected upgoing $\nu_\mu$'s via charged current interactions, \(
\nu _{\mu } \rightarrow \mu \); upgoing muons were identified with the
streamer tube system (for tracking) and the scintillator system (for
time-of-flight measurement).
The events measured and expected for the three measured topologies,
deviate from Monte Carlo expectations without oscillations,
Fig.\ref{fig2}; these deviations and the $L/E_\nu$ distribution point to the
same  ${\nu_\mu \rightarrow  \nu_\tau}$  oscillation scenario
\cite{r2}-\cite{r8}, Fig. \ref{fig3}.
\section{Atmospheric neutrinos. Monte Carlo}
\label{sec:1}
\begin{figure}[th]
\begin{center}
\resizebox{0.7\hsize}{!}{%
  \includegraphics{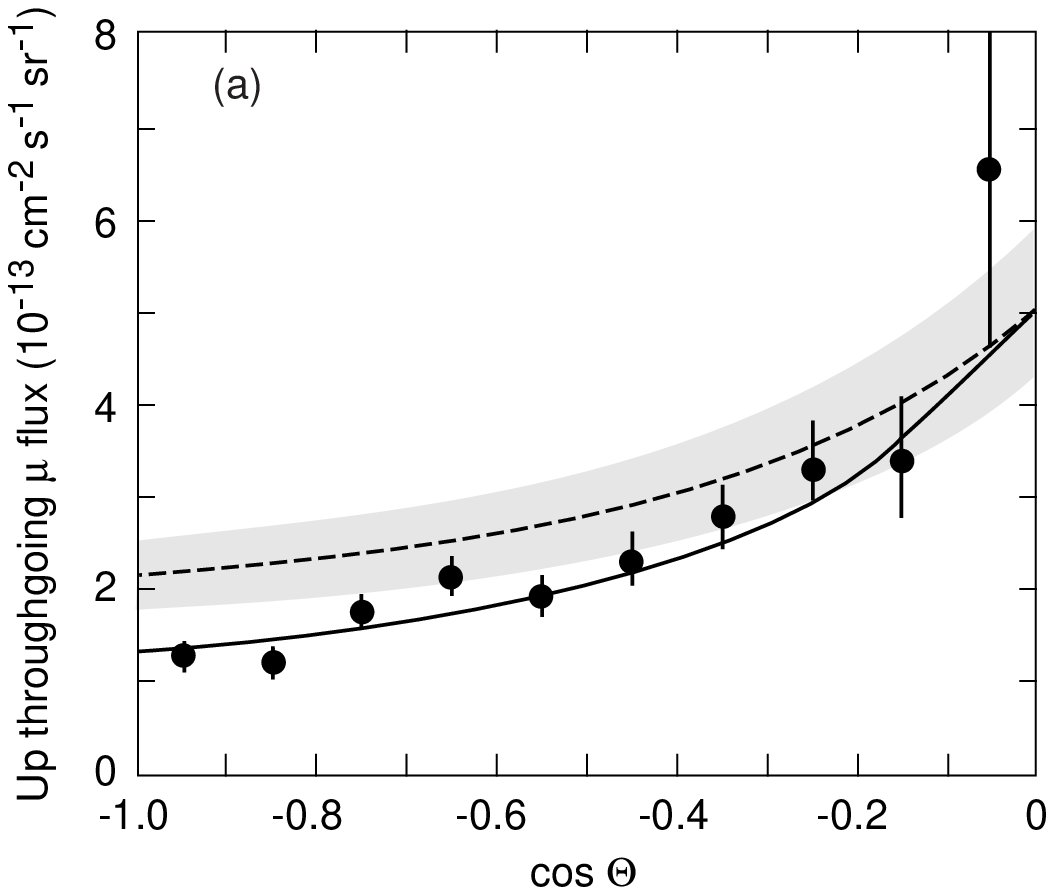} } 
%  \resizebox{0.7\hsize}{!}{%
  \resizebox{0.4\vsize}{!}{%
  \includegraphics{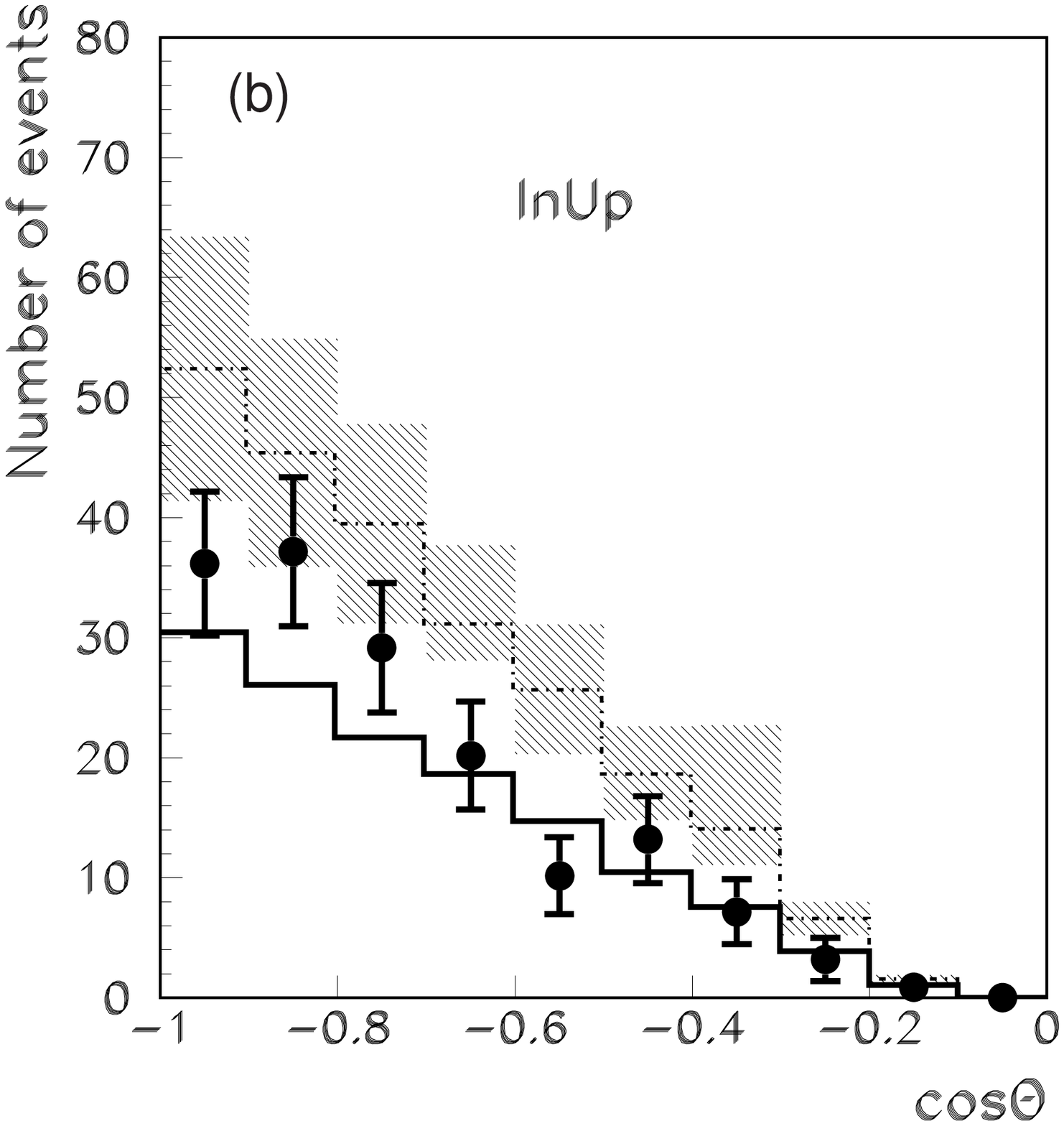}
}
  \resizebox{0.4\vsize}{!}{%
  \includegraphics{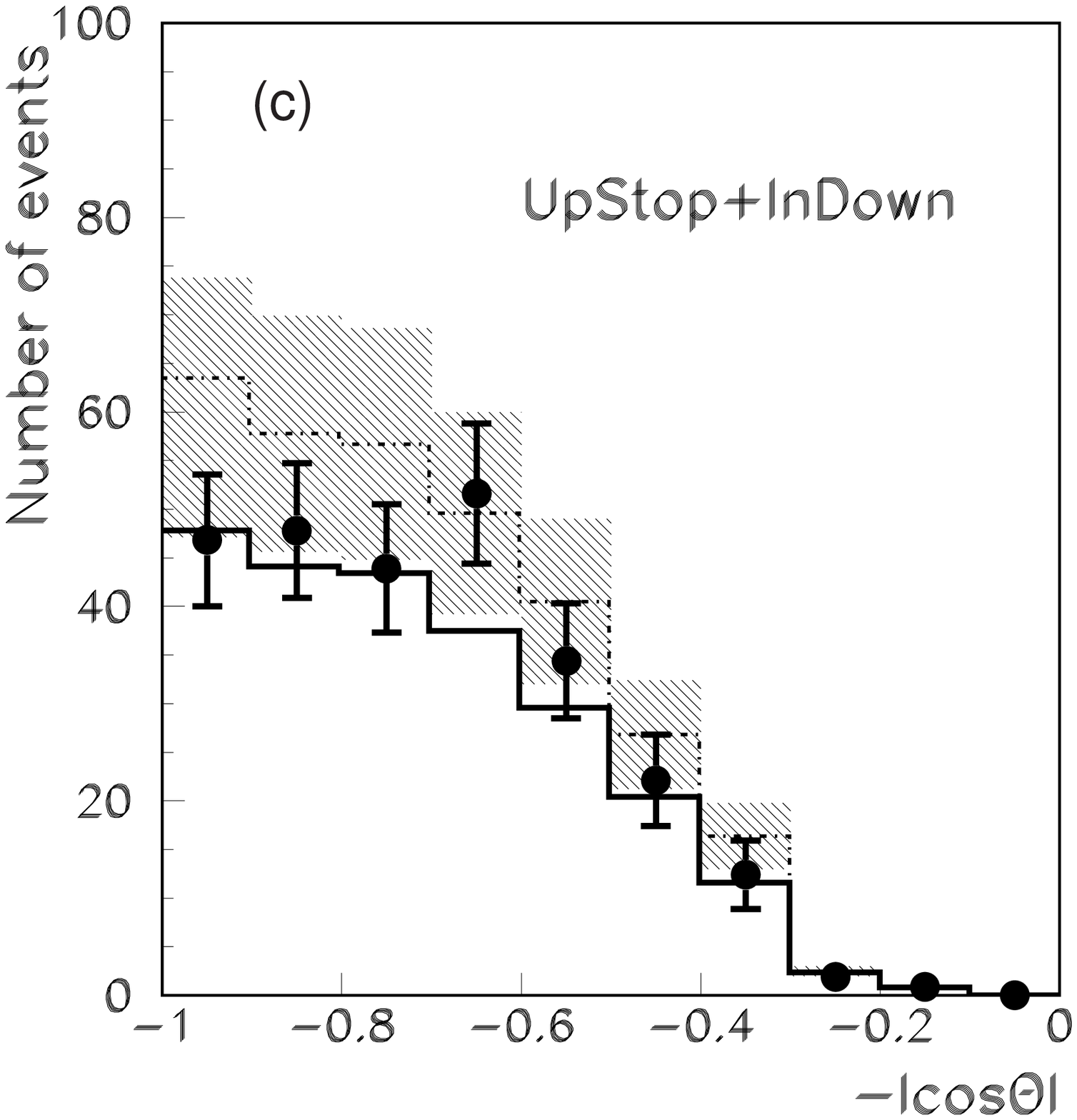}
}
\end{center}
 \caption{\label{fig2}\small Zenith distributions for the MACRO data (black
  points) for (a) upthroughgoing, (b) semicontained and (c) up-stopping
  muons +  down semicontained. The dashed line in (a) is the no-oscillation Bartol96 MC
  prediction (with a scale error band), (in b, c are the FLUKA MC;  better fits are obtained with Bartol96); the solid lines refers to 
 \( \nu _{\mu }\rightarrow  \nu _{\tau } \) oscillations  with maximal
  mixing and $\Delta  m^{2} = 2.3 \cdot 10^{-3}$ eV$^{2}$ (see text).}
\end{figure}

\begin{figure}[th]
\begin{center}
\resizebox{0.9\hsize}{!}{%
  \includegraphics{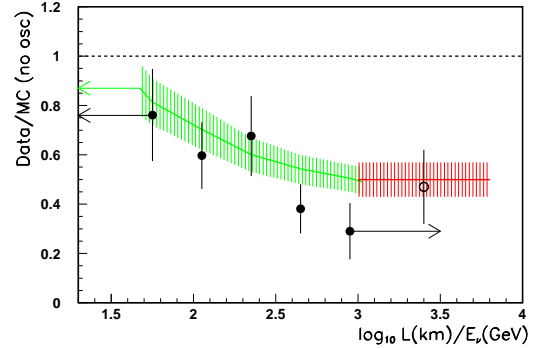}
  }
\end{center}
\caption{\label{fig3}\small  Ratio (Data/MC Bartol96) versus the estimated $L/E_{\nu}$ for the upthrougoing
muon sample  (black circles) and the semicontained up-$\mu$ (open circle). The  horizontal dashed line at
Data/MC=1 is the  expectation for no oscillations.}
\end{figure}

\begin{figure}[th]
\begin{center}
   \resizebox{0.7\hsize}{!}{%
\includegraphics{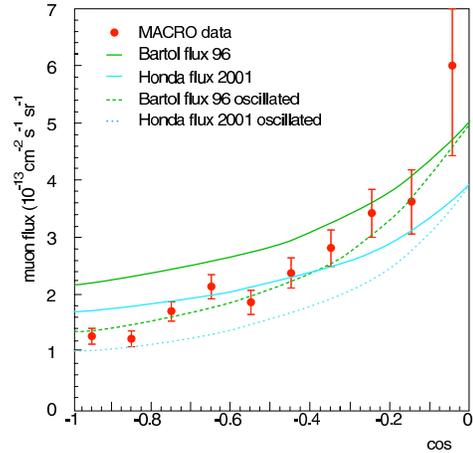}}
\end{center}
\caption{\label{fig3b}\small Comparison of our measurements with  the Bartol96
  and the new Honda 2001 oscillated  and non oscillated fluxes.}
\end{figure}

The measured data of Fig. \ref{fig2} were compared with different MC simulations.
In the past  we used the neutrino flux computed by
the Bartol96 group \cite {bartol} and the GRV94  
parton distribution. For the low energy channels the cross sections by P. Lipari \textit{et al.}
were used; the propagation of muons to the detector used the energy loss calculation by Lohmann \textit{et al.}
The total systematic uncertainty in the predicted flux of
upthroughgoing muons, was estimated at 
$\pm 17 \:\%$;  this is mainly a scale error that does not change the shape
of the angular distribution. The response of the detector to the passage of particles was simulated using GEANT3. A
similar MC (Honda96)  was used by the SuperK Collaboration \cite{hayato} \cite{honda}.\\
Recently new improved MC predictions for neutrino fluxes were made available by the Honda \cite{honda} and FLUKA
\cite {fluka} groups. 
They include three dimensional calculations of hadron production and decays
and of neutrino  interactions, improved hadronic model and new fits of the
primary  cosmic ray flux. The two MC yield predictions for
the non oscillated and oscillated \numu fluxes equal to within few \% \cite{r8}. The shapes of the
angular distributions for oscillated and non oscillated Bartol96, new FLUKA
and new Honda fluxes are the same to within few \%.
The absolute values of our upthroughgoing muon data are about $25 \%$
above those  predicted by the new FLUKA and Honda MC, Fig. \ref{fig3b}. A similar situation is
found in the  new SuperK data \cite{hayato}. 
The high energy \numu data thus suggest that the new Honda and FLUKA predictions are low,
probably because of the used CR fit
(the inclusion of the new ATIC Collab. measurements of primary CRs
may improve the situation \cite{battiston}).
The evidence for neutrino oscillations is due
mainly to the shape  of the angular distribution and this is the same
in all MCs. Also the ratios of the medium to high energy measurements and of the two different samples of low energy data are MC independent. Our low energy data suggest that the FLUKA normalization should be raised by about $12 \%$ at $E_\nu \sim 3 \; GeV$.
\section{MACRO results on atmospheric neutrinos}
The {\it upthroughgoing muons } come from $ \nu _{\mu } $ interactions
in the rock below the detector; muons with \(E_{\mu }>1\GeV  \) cross the whole detector. The corresponding \( \nu _{\mu } \)'s have a median
energy \( \overline{E}_{\nu }\sim \, \, 50\, \, \GeV  \).  Many possible systematic effects and backgrounds
that could affect the measurements were studied \cite{r3,r8}. 
The data, Fig. \ref{fig2}a, deviate in shape and in absolute value from the Bartol96 MC
non oscillated predictions.\\
\textit{\numu $\rightarrow$ \nutau versus \numu $\rightarrow \nu_s$}. 
Matter effects would produce a different total number and a different zenith angle distribution
of upthroughgoing muons.
The ratio $R_1 = $ Vertical/Horizontal $= N(-1<cos\theta < -0.7) / N(-0.4 <
cos\theta < 0)$ was  used to test the \numu $\rightarrow
\nu_s$ oscillation hypothesis versus \numu $\rightarrow$ \nutau \cite{r2} \cite{r6} \cite{r8}. 
The  \numu $\rightarrow \nu_s$ oscillations (with
any mixing) are excluded at about  $99.8\%$ c.l. with respect to \numu
$\rightarrow$ \nutau oscillations with  maximal mixing \cite{r8}.\\
\textit{Oscillation probability as a function of the ratio $L/E_\nu$}. $E_\nu$ was estimated by measuring the muon energy, $E_\mu$, by means of the muon Multiple Coulomb Scattering (MCS) in the rock absorbers in the lower MACRO. The space resolution achieved is \(  \simeq 3{\mm } \). For each muon, seven variables were given in input to
  a Neural Network (NN)  trained to estimate muon energies with  MC events of known input energy crossing the detector at different zenith  angles.   The distribution of the ratio $R = (Data/MC_{no osc})$ obtained
  by this analysis is plotted in Fig. \ref{fig3} as a function of $
  (L/E_\nu)$ \cite{r7}. The data extend from $(L/E_\nu) \sim
  30$ to $ 5000$ km/GeV.
\begin{figure}[th]
\begin{center}
\resizebox{0.4\textwidth}{!}{%
  \includegraphics{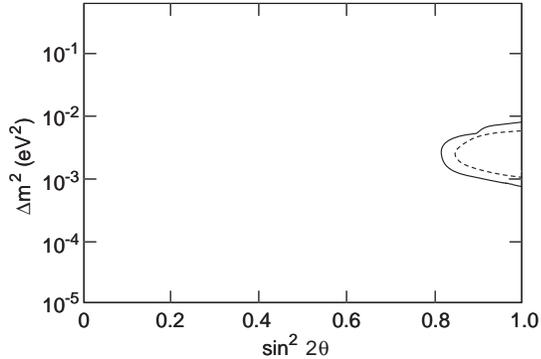} }
\end{center}
\caption{\label{fig5}\small  
Interpolated qualitative 
90\% C.L. contour plots of the allowed regions in the \( \Delta m^{2} -
\sin^{2}2\theta \) plane  for the MACRO data using only the ratios $R_{1},
R_{2}, R_{3}$ (outer continuous line)  and using also the absolute values
assuming the validity of the Bartol96 fluxes (dotted line). 
}
\end{figure}

The \textit{Internal Upgoing} (IU) muons come from  $\sim 3$ GeV $ \nu _{\mu }$'s interacting in the lower
apparatus. Compared to the no-oscillation prediction there is a reduction in
the flux of these events, without distortion in the shape of the zenith
distribution, Fig. \ref{fig2}b. The MC predictions for no oscillations in
Figs. \ref{fig2}b and \ref{fig2}c are  the dashed lines with a $21
\:\%$ systematic   band. At these energies the ratios DATA/MC are 0.67 for FLUKA and 0.54 for Bartol96.

The \textit{Upstopping} (UGS) muons are due to $\sim 3$ GeV  \( \nu _{\mu } \)'s
interacting below the detector, yielding upgoing muons stopping in the
detector. The \textit{Semicontained Downgoing } (ID) muons are due to \( \nu
_{\mu } \)-induced downgoing $\mu$'s with vertex in the lower MACRO. 
The two types of events are identified by 
topological criteria. The upgoing $\nu_{\mu}$'s should  have oscillated completely, while the downgoing \numu do not.
\section{Determination of the oscillation parameters.}
\label{sec:2}
In the past, in order to determine the oscillation parameters, we made fits
to the shape of the upthroughgoing muon zenith distribution and to the absolute
flux compared to the Bartol96 MC prediction. The other data were only used to verify the consistency
and to make checks. The result  was $\Delta m^{2} = 0.0025$ eV$^2$ and
maximal mixing \cite{r6} \cite{r3}. Later, also the $L/E_\nu$ distribution was considered \cite{r7}.

In order to reduce the effects of possible systematic uncertainties in the MC we recently used the following three independent ratios \cite{r8} and we checked that FLUKA, Honda and Bartol96 Monte Carlo simulations yield the same predictions to within $\sim 5 \%$.
\begin{enumerate}
        \item [(i)] High Energy Data: zenith distribution ratio: $R_{1} = N_{vert}/N_{hor}$
        \item [(ii)] High Energy Data, neutrino energy measurement ratio: $R_{2} = N_{low}/N_{high}$
        \item [(iii)] Low Energy Data: \\Ratio $R_{3} = (Data/MC)_{IU}/(Data/MC)_{ID+UGS}$.
\end{enumerate}
With these ratios, the no oscillation hypothesis has a probability $P \sim 3 \cdot 10^{-7} $
and is thus ruled out by  $ \sim 5 \sigma$.
%The formula used for  combining independent probabilities is $P =
%P_{1}P_{2}P_{3} (1  - \ln P_{1}P_{2}P_{3} + 1/2 (\ln P_{1}P_{2}P_{3})^2)$\cite{roe}. 
By fitting the three ratios  to the \numu $\rightarrow$ \nutau oscillation
formulae we  obtain $\sin^{2} 2\vartheta = 1$,   $\Delta  m^{2} = 2.3
\cdot 10^{-3}$ eV$^{2}$ and the allowed region indicated by the solid line in Fig. \ref{fig5}. 

There is a good consistency between the old and new methods. 

If we use the Bartol96 flux we may  add to the ratios (i) - (iii)  the information on the absolute flux values of the
\begin{enumerate}
   \item[(iv)] high energy data (systematic scale error of $\gsim 17 \%$) $R_{4} =
     N_{meas}/N_{MCBartol}$.
   \item [(v)] low energy semicontained muons, with a systematic scale
     error of $21 \%$, $R_{5} \simeq N_{meas}/N_{MCBartol}$.
     \end{enumerate}
These informations reduce  the area of the allowed region in the 
\( \Delta m^{2} - \sin^{2}2\theta \) plane, as indicated by the dashed line
in Fig. \ref{fig5}.
The limit lines represent smoothed interpolations and are qualitative. 
The final MACRO $\Delta  m^{2}$ is $ 2.3 \cdot 10^{-3}$ eV$^{2}$. \\
%Our data suggest that the new FLUKA and Honda MC predictions for the $\nu$ flux should 
%be raised by $\sim 25 \%$ at $\bar E_{\nu}\sim 50$ GeV and $\sim 12 \%$ at 
%$\bar E_\nu \sim 3$ GeV.\\
We would like to acknowledge the cooperation of  the members of the MACRO collaboration. 

% BibTeX users please use
% \bibliographystyle{}
% \bibliography{}

\end{document}